\documentclass[aps,prb,twocolumn,sd,amsmath,amssymb,final,superscriptaddress]{revtex4-1}
\usepackage{dcolumn}
\usepackage{graphicx}
\usepackage{epstopdf}
\usepackage{amsmath}
\usepackage[bookmarks=false,pdfstartview=FitH]{hyperref}
\usepackage{color}
\usepackage{booktabs}
\usepackage[all]{hypcap}
\usepackage{ragged2e}

\newcommand\T{\rule{0pt}{2.6ex}}       
\newcommand\B{\rule[-1.2ex]{0pt}{0pt}} 

\definecolor{orange}{cmyk}{0,0.4,0.8,0.2}
\definecolor{darkorange}{rgb}{.71,0.21,0.01}
\definecolor{darkgreen}{rgb}{.12,.54,.11}
\definecolor{darkblue}{rgb}{0.1,0.1,0.8}
\hypersetup{pdftex,
  breaklinks=true,  
  colorlinks=true,
  urlcolor=blue,
  linkcolor=blue,
  citecolor=blue,
  }

\def\Let@{\def\\{\notag\math@cr}}

\begin{document}


\title{Self-compensation in phosphorus-doped CdTe}


\author{Mauricio A. Flores}
\email[]{mauricio.flores@ug.uchile.cl}
\affiliation{Facultad de Ingenier\'ia y Tecnolog\'ia, Universidad San Sebasti\'an, Bellavista 7, Santiago 8420524, Chile.}

\author{Walter Orellana}
\affiliation{Departamento de Ciencias F\'isicas, Universidad Andres Bello, Sazi\'e 2212, 037-0136 Santiago, Chile.}

\author{Eduardo Men\'endez-Proupin}
\affiliation{Departamento de F\'isica, Facultad de Ciencias, Universidad de Chile, Las Palmeras 3425, 780-0003 \~Nu\~noa, Santiago, Chile.}

\begin{abstract}
We investigate the self-compensation mechanism in phosphorus-doped CdTe. The formation energies, charge transition levels, and defects states of several P-related point defects susceptible to cause self-compensation are addressed by first-principles calculations.
Moreover, we assess the influence of the spin-orbit coupling and supercell-size effects on the stability of AX centers donors, which are believed to be responsible for most of the self-compensation. We report an improved result for the lowest-energy configuration of the P interstitial $(\text{P}_\text{i})$ and find that the self-compensation mechanism is not due to the
formation of AX centers. Under Te-rich growth conditions, $(\text{P}_\text{i})$ exhibits a formation energy lower than the substitutional acceptor $(\text{P}_\text{Te})$ when the Fermi level is near the valence band, acting as compensating donor. While, for Cd-rich growth conditions, our results suggest that \emph{p-}type doping is limited by the formation of $(\text{P}_\text{Te}-\text{V}_\text{Te})$ complexes.
\end{abstract}

\date{\today}


\maketitle
\section{Introduction}

Cadmium telluride (CdTe) is one of the few II-IV semiconductors that can be doped both \emph{n-} and \emph{p-}type.\cite{Chen14} Due to its near optimal band gap of 1.5 eV and high absorption coefficient near the band edge, CdTe is used as a semiconductor absorber layer in photovoltaic technology. However, CdTe-based solar cells usually exhibit an open-circuit voltage (V$_\text{oc}$) of $\sim$\hspace{0.05cm}860 mV, which is low in comparison to the detailed-balance limit of 1.23 V. \cite{Zhao16} One route to improve the V$_\text{oc}$ is by increasing both hole density and carrier lifetime. \cite{Gessert13,Duenow16,Yang16} The latter can be improved by reducing the concentration of defects that introduce levels deep in the band gap, which act as effective non-radiative Shockley-Read-Hall (SRH) recombination centers. \cite{Flores16_2,Wickramaratne16} On the other hand, hole density can be boosted by the incorporation of shallow acceptors. These are effective \emph{p-}type dopants because they can be easily ionized at room temperature, providing free holes to the valence-band. \cite{Lyons14} In practice, however, \emph{p-}type doping is difficult to achieve in CdTe, and this has so far hindered the production of high hole density films.\cite{Krantz13} The origin of the low \emph{p-}type dopability in CdTe is not yet well understood, but it has been suggested that it may be due to the low solubility of the dopants, the introduction of deep levels, or self-compensation. \cite{Korcsmaros16,Ablekim17}
The latter term denotes the response of the system to the introduction of electrically active impurities, which tends to compensate their electrical activity through the formation of opposite charged defects. \cite{Mandel64,Krasikov13}

To improve the \emph{p}-type conductivity in CdTe, extrinsic doping with group-I (Li, Na, Cu, Ag) and group-V elements (N, P, As, Sb, and Bi) has been explored in the literature. Group-I elements are expected to occupy the Cd site introducing shallow acceptors levels, but \emph{p-}type conductivity is usually limited by self-compensation. Moreover, it has been reported that Li, Na, and Ag exhibit an amphoteric behavior acting as substitutional acceptors and interstitial donors; \cite{Desnica98,Wei02} whereas Cu exhibits a fast diffusion mechanism that leads to the formation of complexes with Cd vacancies, deteriorating the substitutional acceptor states.\cite{Jones92,Grecua00,Dzhafarov05,Yang17} Group-V elements, on the other hand, are natural substitutes for tellurium and are also expected to introduce shallow acceptor levels in CdTe. Recent experimental works suggest that doping with group-V elements under Cd-rich conditions can indeed increase the hole concentrations, \cite{Burst16, Ablekim17, Nagaoka17} but also reveal the existence of an apparent doping limit. \cite{Nagaoka17} The origin of this limiting mechanism is not well understood, but recent theoretical studies suggest that it is due to the formation of self-compensating positive charged AX centers. \cite{Yang15,Colegrove16,Colegrove17,Ablekim17}

In this work, we investigate the origin the \emph{p}-type doping difficulties in phosphorus-doped CdTe. We present improved results for the lowest-energy interstitial configuration $(\text{P}_\text{i})^{+1}$ reported by Colegrove et al., \cite{Colegrove16} and find that the self-compensation mechanism is not due to the formation of AX centers. Under Cd-rich growth conditions, the self-compensation occurs through the formation of phosphorus complexes with tellurium vacancies, whereas under Te-rich growth conditions the $(\text{P}_\text{i})^{+1}$ configuration plays the compensating role, limiting the \emph{p-}type conductivity.

\section{Methods}

\subsection{Computational details}

We performed first-principles DFT calculations within the generalized gradient approximation (GGA) formulated by Perdew, Burke, and Ernzerhof (PBE),\cite{Perdew96} as implemented in the Quantum-ESPRESSO code. \cite{Giannozzi2009} Unless otherwise stated, our total-energy and band-structure calculations were performed employing GBRV ultrasoft pseudopotentials \cite{Garrity2014} with a plane-wave energy cutoff of 36 Ry. The defect formation energies were calculated using 512-atom supercells, in which the sampling of the Brillouin zone was restricted to the $\Gamma$ point. In the case of positive and negative charged systems, we applied a correction of 0.08 eV which results from the application of scheme proposed by Lany and Zunger. \cite{Lany08} In all calculations, the structures were relaxed until the Hellmann-Feynman forces on each atom were less than 0.001 Ry/bohr.

Additionally, we performed \textit{GW} calculations in 64-atom supercells to obtain the quasiparticle corrections to the Kohn-Sham band structure in selected cases. These corrections were considered through the application of scissors operators at the $\Gamma$ point on larger 512-atom supercells. We used the DFT+U approach \cite{Kioupakis08} as starting point for a subsequent COHSEX+$G_0W_0$ calculation using the ABINIT code.\cite{Gonze2009,Gonze16} A $2\times{2}\times{2}$ \textbf{k}-mesh was used to obtain the converged DFT charge density and the $\Gamma$-point only in the subsequent \textit{GW} calculation. We used a 20 Ry energy cutoff to represent the dielectric matrix $(\epsilon)$ and 2048 bands plus the extrapolar approximation of Bruneval and Gonze. \cite{Bruneval08} The frequency dependence of the dielectric matrix was approximated by the plasmon-pole model of Godby and Needs, \cite{Godby89} which requires that the behavior of $\epsilon^{-1}$ is correctly reproduced at two different frequencies: the static limit ($\omega = 0$) and an additional imaginary frequency near the plasma frequency of the system. The Hubbard parameter was set to \emph{U}$_\text{Cd}= 7 $ eV for the 3$d$ states of Cd.

\subsection{Defect formation energies}

The formation energy of a given defect or impurity determines its concentration. \cite{Freysoldt14} The defect formation energy in charge state $q$ and arbitrary ionic configuration can be expressed as \cite{Jain11, Flores16_1, Flores16_2,Flores17_2}

\begin{eqnarray}
E^f_q[\textbf{R}] = E_q[\textbf{R}] - E_\text{ref} + qE_F,
\label{ec:1}
\end{eqnarray}
\begin{eqnarray}
E_\text{ref} \equiv E^\text{CdTe}_\text{bulk} + \sum_i n_i(\Delta\mu_i + \mu^\text{ref}_i),
\end{eqnarray}
where $E_q[\textbf{R}]$ is the total energy of the system in charge state $q$ and ionic configuration $\textbf{R}$, $E_\text{ref}$ is the energy of a reference system, i.e. an equivalent defect-free supercell, and $E_F$ corresponds to the Fermi level. The integer $n_i$ represents the number of atoms of species $i$ that are either added ($n_i > 0$) or removed ($n_i < 0$) from the reference supercell. $\Delta\mu_i$ is the relative chemical potential for the $i$th atomic species referenced to $\mu^\text{ref}_i$, which is the chemical potential of its pure elemental phase, e.g., Cd (hexagonal structure with space group $P6_3$), Te (trigonal structure with space group $P3_121$), and P (orthorhombic structure with $Cmca$ space group).

Additionally, upper and lower bounds to the chemical potentials are required to maintain the stability of the CdTe compound:
\begin{equation} \Delta\mu_{\text{Cd}} + \Delta\mu_{\text{Te}} = E^f[\text{CdTe}], \end{equation}
where $E^f[\text{CdTe}] = -0.91$ eV is the calculated formation enthalpy of CdTe. Moreover, the upper bound of $\Delta\mu_{\text{Cd}}$ (and the lower bound of $\Delta\mu_{\text{Te}}$) corresponds to the reduction of CdTe to metallic Cd, i.e. $\Delta\mu_{\text{Cd}} = 0$, whereas the lower bound of $\Delta\mu_{\text{Cd}}$ (and thus the upper bound of $\Delta\mu_{\text{Te}}$) is given by the reduction of CdTe to elemental Te, i.e. $\Delta\mu_{\text{Cd}} = -0.91$ eV. To avoid the formation of secondary phases of P with the host atoms, its chemical potential is bounded by the following relation:
\begin{equation} 3\Delta\mu_{\text{Cd}} + 2\Delta\mu_{\text{P}} \leq E^f[\text{Cd}_3\text{P}_2] = -0.35 \text{ eV}, \end{equation}
where $E^f[\text{Cd}_3\text{P}_2]$ is the calculated formation enthalpy of $\text{Cd}_3\text{P}_2$ (tetragonal structure with space group $P4_2 / nmc$). Moreover, to prevent the precipitation of P, its chemical potential should also be smaller than of the corresponding elemental phase. Thus $\Delta\mu_{\text{P}}$ is restricted by:
\begin{equation} \Delta\mu_{\text{P}} \leq \min \left(0,\frac{1}{2}(E^f[\text{Cd}_3\text{P}_2]- 3\Delta\mu_{\text{Cd}})\right).\end{equation}
For a Cd-rich growth condition we have $\Delta\mu_{\text{Te}} = -0.91$ eV and $\Delta\mu_{\text{Cd}} = 0$ eV, then
\begin{equation} \Delta\mu_{\text{P}}^\text{Cd-rich} = -0.18 \text{ eV}, \end{equation}
and, for a Te-rich growth condition we have $\Delta\mu_{\text{Te}} = 0$ eV and $\Delta\mu_{\text{Cd}} = -0.91$ eV, thus
\begin{equation} \Delta\mu_{\text{P}}^\text{Te-rich} = 0 \text{ eV}. \end{equation}

\begin{figure}[h]%
\vspace{0.2cm}
 \centering
 \includegraphics[width=3.6cm]{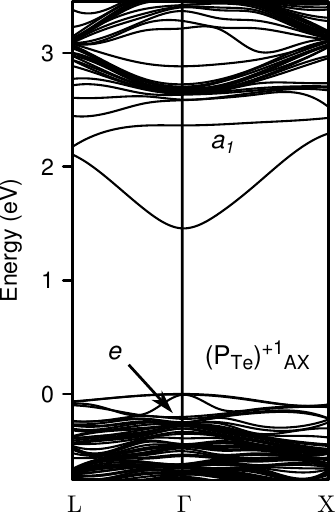}
  \includegraphics[width=4.8cm]{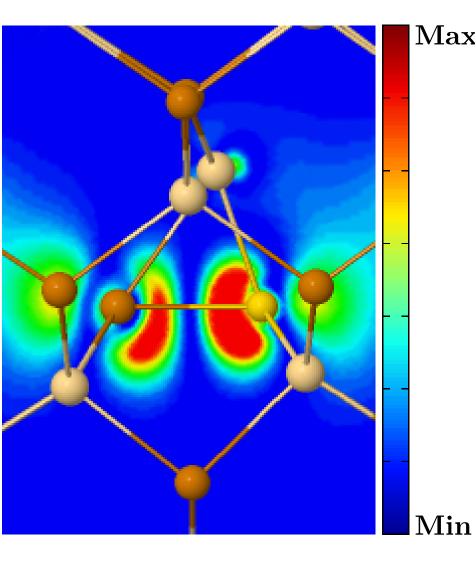}
 \caption{(Left) Electronic band structure of $(\text{P}_\text{Te})^{+1}$ in the AX configuration. (Right) Squared wave function corresponding to the anti-bonding $a_1$ level plotted in the range of $0-4\times10 ^{-3}$ bohr$^{-3}$. Dark and light spheres represent Te and Cd atoms, respectively, the yellow sphere represents the phosphorus impurity. The calculations were performed in a 250-atom supercell.}
 \label{fig1}
  \vspace{0.3cm}
\end{figure}

\section{Results and discussion}

\subsection{Stability of AX centers}

In an AX center a group-V substitutional dopant (occupying a Te site) and its nearest neighbour Te atom move toward each other, breaking their original bonds with the Cd atoms along the [110] zigzag chain to form a new bond. \cite{Zhang02} This ionic distortion results in the net loss of one bond and in the creation of an anti-bonding orbital above the conduction-band minimum (CBM), as shown in Figure \hyperref[fig1]{1}. Moreover, two electrons are released to the Fermi sea transforming the negative charged substitutional acceptor into a positive charged donor.

 \begin{table}[h]
\vspace{-0.15cm}
\centering
\caption{\label{tab:t1} Energy difference between T$_\emph{d}$ and AX configurations in the positive charge state, for several supercell sizes (all values are given in eV). }
\begin{ruledtabular}
\begin{tabular}{cccc}
 \hspace{0.5cm}\text{Size} & \hspace{0.3cm}\textbf{k}-point sampling & \hspace{0.1cm}functional & E(${\text T_d})-$E(AX) \B \\
\hline
\T  \T \hspace{0.5cm}64 & \hspace{0.15cm}$2\times{2}\times{2}$ & PBE & \hspace{-0.2cm}$-0.25$\\
\T \hspace{0.5cm}64 & \hspace{0.15cm}$2\times{2}\times{2}$ & HSE06 & \hspace{0.2cm}$0.50^{a}$\\
\T  \hspace{0.5cm}216 & \hspace{0.15cm}$\Gamma$ & PBE & \hspace{-0.2cm}$-0.56$\\
\T  \hspace{0.5cm}216 & \hspace{0.15cm}$2\times{2}\times{2}$ & PBE & \hspace{-0.2cm}$-0.47$\\
\T  \hspace{0.5cm}216 & \hspace{0.15cm}$\Gamma$ & HSE06 & \hspace{0.05cm}$0.32$\\
\T \hspace{0.5cm}512 & \hspace{0.15cm}$\Gamma$& PBE & \hspace{-0.1cm}$-0.62$ \vspace{0.2cm}\\
\end{tabular}
\begin{flushleft}
\footnotesize\label{a}$^{a}$Ref. [\onlinecite{Yang15}]
\end{flushleft}
\end{ruledtabular}
\end{table}

Further insights can be gained from simple valence arguments. The neutral tellurium vacancy (V$_\text{Te}$) in T$_\emph{d}$ symmetry introduces a fully occupied $a_1$ level with $s$-like symmetry and an empty triple-degenerate $t_2$ level with \emph{p}-like symmetry. \cite{Lany01} The five valence electrons of the group-V dopant substituting a Te atom only partially fill the $t_2$ level left by the vacancy, and thus one might expect an orbitally-degenerate configuration. However, the $t_2$ level is resonant with the valence-band. Hence, the hole introduced by the dopant occupies a “perturbed host state” (PHS) at the valence-band maximum (VBM), weakly bounded to the impurity in an effective-mass acceptor state. Due to finite-size effects, this PHS merges with the valence-band at typical supercell sizes. \cite{Lany10}

 \begin{table}[h]
\caption{\label{tab:t2} Energy difference between T$_\emph{d}$ and AX configurations in the positive charge state (all values are given in eV). The calculations were performed in a 64-atom supercell, using PAW pseudopotentials \cite{Bloch94} from the PS Library 0.3.1, \cite{DalCorso14} 40 Ry plane-wave energy cutoff, and a $2\times2\times2$ Monkhorst-Pack \textbf{k}-point grid. SR and FR refer to scalar-relativistic (neglecting SOC) and fully-relativistic (including SOC) calculations, respectively. }
\begin{ruledtabular}
\vspace{0.1cm}
\begin{tabular}{ccc}
 \hspace{0.5cm}\text{Dopant} & \hspace{0.4cm}E(${\text T_d})-$E(AX) (SR) & E(${\text T_d})-$E(AX) (FR) \B \\
\hline
\T  \T \hspace{0.5cm}N & \hspace{0.15cm}$-0.39$ & \hspace{-0.2cm}$-0.83$\\
\T \hspace{0.5cm}P & \hspace{0.15cm}$-0.25$  & \hspace{-0.2cm}$-0.67$\\
\T  \hspace{0.5cm}As & \hspace{0.15cm}$-0.21$ & \hspace{-0.2cm}$-0.62$\\
\T  \hspace{0.5cm}Sb & \hspace{0.15cm}$-0.18$  & \hspace{-0.2cm}$-0.55$\\
\T \hspace{0.5cm}Bi & \hspace{0.15cm}$-0.15$ & \hspace{-0.2cm}$-0.60$  \\
\end{tabular}
\end{ruledtabular}
\vspace{0.1cm}
\end{table}

Interestingly, the ionic distortion from the substitutional acceptor $(\text{P}_\text{Te})$ in T$_\emph{d}$ symmetry to the less symmetrical AX center splits the $t_2$ manifold (which is fully occupied) into a filled $e$ state below the VBM and an empty $a_1$ level resonant with the conduction-band, as shown in Figure  \hyperref[fig1]{1}. We should note that the AX center has all the valence bands full and all the conduction bands empty. This is important if we want to compare the total energies of the T$_d$ and the AX configuration in the same charge state, because a finite-size correction due to the delocalized holes occupying the PHS should be applied to the former configuration. To illustrate this, we calculate their total energy differences, E(${\text T_\emph{d}})-$E(AX), in the positive charge state at different supercell sizes as shown in Table \hyperref[tab:t1]{I}. At scalar relativistic DFT-PBE level (spin-orbit free), the T$_\emph{d}$ configuration is more stable than the AX center by 0.25, 0.47, and 0.62 eV by using 64, 216, and 512 supercell sizes, respectively, giving rise to a finite-size correction of 0.37 eV obtained by comparing the 64-atom supercell with the larger 512-atom supercell. Recently, Yang et al. \cite{Yang15} reported that the AX is more stable than the T$_d$ configuration by $0.50$ eV when the HSE06\cite{Heyd06} hybrid functional is used. This effect can be attributed to the downshift in the absolute position of the VBM due to the inclusion of a fraction of Hatree-Fock interaction, lowering the total energy of the AX configuration to a greater extend than the energy of the T$_d$ configuration (since the former has no holes in the valence-band). However, the calculations in Ref. [\onlinecite{Yang15}] were carried on a 64-atom supercell; if larger supercells are employed, finite-size effects will stabilize the T$_\emph{d}$ configuration. In particular, if a large supercell is used the AX center is expected to be stable by only 0.13 eV.

In Table \hyperref[tab:t1]{I}, we present our results for several supercell sizes. If we compare the total energy differences between $(\text{P}_\text{Te})^{+1}_{\text{Td}}$ and $(\text{P}_\text{Te})^{+1}_{\text{AX}}$ obtained with 64-atom (2x2x2 \textbf{k}-mesh) and 216-atom ($\Gamma$-only) supercells, we find a finite size correction of 0.31 eV and 0.18 eV for PBE and HSE06, respectively; favoring the T$_d$ configuration. The slower convergence of the hybrid functional arises from the long-range nature of the Hartree-Fock (HF) exchange interaction, which converges slowly with respect to the supercell size. \cite{Broqvist09, Bang13} Thus, the use of large supercells and/or a careful choice of the screening parameter $(\mu)$, which defines the extent of the HF exchange in real space, is recommended.

\begin{figure}[h]%
 \centering
 \includegraphics[width=3.6cm]{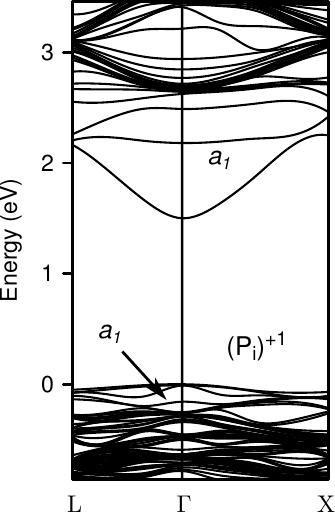}
  \includegraphics[width=4.8cm]{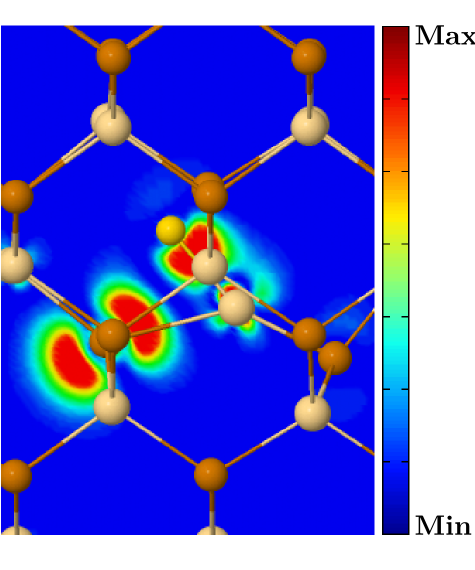}
 \caption{(Left) Electronic band structure of $(\text{P}_\text{i})^{+1}$. (Right) Squared wave function corresponding to the $a_1$ level below the VBM, plotted in the range of $0-4\times10 ^{-3}$ bohr$^{-3}$. Dark and light spheres represent Te and Cd atoms, respectively, the yellow sphere represents the phosphorus impurity. The calculations were performed in a 250-atom supercell.}\label{fig2}
  \vspace{0.3cm}
\end{figure}

\subsection{The effect of the spin-orbit coupling}

 In addition, there is another effect (so far neglected in previous works) that may influence the stability of the AX centers: the spin-orbit coupling (SOC), that affects the absolute position of the VBM and splits the six \textit{p}-shell of the P impurity into two $p_{1/2}$ spinors and four $p_{3/2}$ spinors.
 In the case of P, the dominant effect is the upshift of the absolute position of the VBM by $0.3$ eV, thereby stabilizing the T$_d$ configuration against the formation of AX centers.
 To verify this, we compared the total energies of these two configurations in a 64-atom supercell using the HSE06 functional with the default exchange parameters ($\alpha = 0.25$)  as implemented in the Vienna Ab Initio Simulation Package (VASP),\cite{Kresse96} and using a $2\times2\times2$ Monkhorst-Pack \textbf{k}-point grid. We find that when the SOC is included the AX center is more stable than the T$_d$ configuration by $0.1$ eV. We have done the same calculations for several group-V dopants in CdTe, both at scalar relativistic (spin-orbit free) and fully relativistic (including spin-orbit) DFT level, but employing the PBE functional instead of HSE06. Our results, neglecting finite-size corrections, are shown in the Table I. We find that in all cases the SOC stabilizes the T$_\emph{d}$ configuration; however, as we move down the group-V column, the AX center becomes progressively more stable with the sole exception of bismuth.

It is interesting to note that in the case of P the effect of the SOC is roughly independent of the functional employed. The SOC destabilizes the AX center by 0.42 eV and 0.40 eV, in the case of PBE and HSE06, respectively.
 Moreover, our calculations suggest that the stability of the AX centers strongly depends on the absolute position of the VBM, which has been recognized as a critical quantity for determining defect formation energies and charge transition levels in semiconductors and insulators. \cite{Freysoldt16,Flores16_2}
For the case of P-doped CdTe, our results reveal that the AX center becomes unstable by $0.27$ eV when finite-size effects are accounted for by using the HSE06 hybrid functional + SOC.

\subsection{Comparison with the DFT+\textit{GW} approach}

Next, we calculate the $\epsilon (+/-)$ charge transition level induced by the positive charged AX center and the substitutional (P$_\text{Te})^{-1}$ in T$_d$ symmetry by using the DFT+\emph{GW}approximation, as described in Refs. [\onlinecite{Flores16_1,Flores16_2}]. We applied the DFT+\emph{GW} scheme starting from the (P$_\text{Te})^{-1}$  configuration, as the self-interaction error mostly cancels in the first difference of Eq.  (\hyperref[ec:1]{1}). The position of the $\epsilon (+/-)$ is found at VBM + $0.31$ eV; thus, the AX center is stable when the $E_F$ is lower than this value. The large discrepancy between DFT+\emph{GW} and HSE06 could be explained by the fact that hybrid functionals are still affected by the self-interaction error which artificially raises the position of the VBM. This can be seen by comparing experimental results for ionization potentials of several semiconductors, with those obtained from one-electron Green's
function methods and hybrid functionals, as shown in Figure 1 of Ref. [\onlinecite{Gruneis14}]. In addition, it should be noted that a HSE06 calculation using a 64-atom supercell, but neglecting both finite-size and SOC effects results in a fortuitous error cancellation that leads to the stabilization of the AX centers.

\subsection{Self-compensation model for P-doped CdTe}

We have made an extensive search for compensating defects other than the AX center that could potentially introduce charge transition levels in the gap.
We identified an interstitial configuration $(\text{P}_\text{i})^{+1}$ (shown in Figure \hyperref[fig2]{2}), which is $0.7$ eV lower in energy than the split configuration reported in Ref. [\onlinecite{Colegrove16}] (by comparing their total energies in a 64-atom supercell and using the HSE06 hybrid functional).
The split configuration in the positive charge state has two localized levels in the band gap, one of which fully occupied.
In the $(\text{P}_\text{i})^{+1}$ configuration, however, these levels are split across the band gap, giving rise to a ground-state configuration with all the valence bands filled and all the conduction bands empty.
The electronic structure of the $(\text{P}_\text{i})^{+1}$ configuration and the squared wave function corresponding to the $a_1$ level resonant with the valence-band are shown in Figure \hyperref[fig2]{2}. Furthermore, we calculate the diffusion barrier between two adjacent $(\text{P}_\text{i})^{+1}$ configurations using the nudged elastic band (NEB) \cite{Mills94} method, as implemented in the Quantum-Espresso code, \cite{Giannozzi2009} and find a small barrier of 0.61 eV.

\begin{figure}[h]%
\vspace{0.3cm}
 \centering
  \includegraphics[width=8.7cm]{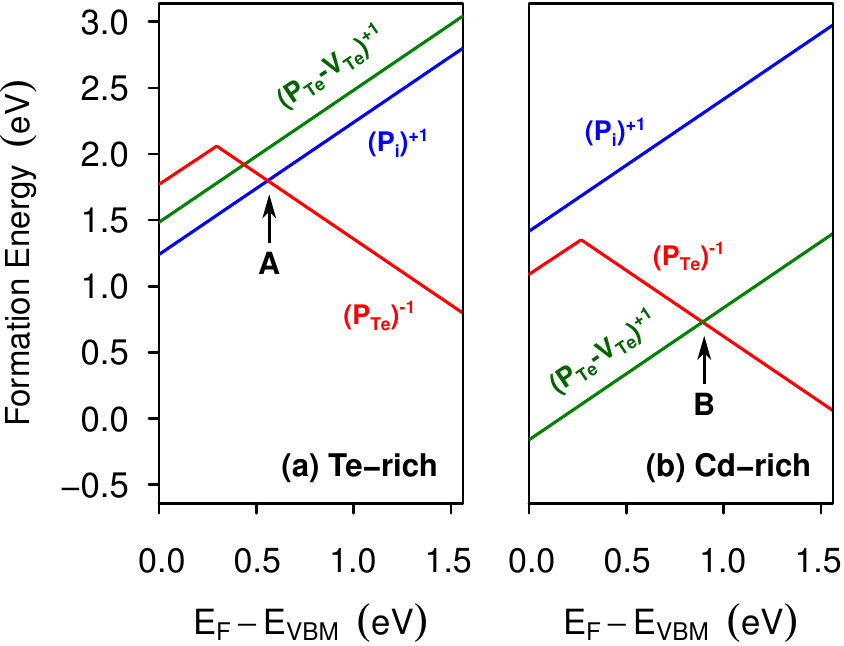}
 \caption{Calculated defect formation energies of $(\text{P}_\text{Te})^{-1}$, $(\text{P}_\text{i})^{+1}$, and $(\text{P}_\text{Te}-\text{V}_\text{Te})^{+1}$  as a function of the Fermi level inside the band gap under (a) Te-rich condition and (b) Cd-rich condition. The formation energy of the AX configuration obtained from the DFT+$GW$ scheme is also included.}\label{fig3}
\end{figure}

 We also investigate the possible formation of defect complexes of P with Cd or Te vacancies. We find that $(\text{P}_\text{Cd}-\text{V}_\text{Cd})$ , $(\text{P}_\text{Te}-\text{V}_\text{Cd})$, and $(\text{P}_\text{Cd}-\text{V}_\text{Te})$ complexes have high formation energies and are unlikely to form at substantial concentrations.
The $(\text{P}_\text{Te}-\text{V}_\text{Te})^{+1}$ complex, on the contrary, exhibits a low formation energy being stable only in the positive charge state. The calculated formation energies for the $(\text{P}_\text{Te}-\text{V}_\text{Te})^{+1}$ complex, the substitutional acceptor $(\text{P}_\text{Te})^{-1}$, and the interstitial configuration $(\text{P}_\text{i})^{+1}$ are plotted in Figure \hyperref[fig3]{3}.
It should be noted that all of these defects are in ground-state configurations, i.e. all the valence bands are full and all the conduction bands are empty; thus, the self-interaction and band gap error on the formation energy is expected to be small at DFT level, only a correction in the absolute position of the VBM is required ($\Delta E_\text{VBM} = -0.74$ eV in the case of CdTe) \cite{Flores16_1, Flores16_2} to consider it as a proper reference for the Fermi level.

Under Te-rich growth conditions, for values of $E_F$ higher than VBM\hspace{0.1cm}+\hspace{0.1cm}0.56 eV (point A in Figure  \hyperref[fig3]3), we find that the most stable configuration is the substitutional acceptor $(\text{P}_\text{Te})^{-1}$. On the other hand, when $E_F$ is close to the VBM, the formation of P interstitials in the positive charge state is favored. Under this condition, both defects tend to compensate each other leading to the Fermi-level pinning close the point A. Unfortunately, Te-rich conditions also favor the formation of Te antisites, which act as hole traps with a deleterious
impact on carrier transport. \cite{Flores16_2} For Cd-rich growth conditions, we find that for values of $E_F$ higher than VBM\hspace{0.1cm}+\hspace{0.1cm}0.89 eV (point B in Figure \hyperref[fig3]3) the substitutional acceptor $(\text{P}_\text{Te})^{-1}$ is favored; whereas, for lower values of $E_F$, the P impurity tends to form vacancy-impurity complexes with Te vacancies being a limiting factor for an efficient \emph{p-}type doping. Under this growth condition, the Fermi-level might be pinned near point B. However, it is important to keep in mind that this situation represents a limit case in which the partial pressure of Cd is such that P impurities are in the boundary of forming a $\text{Cd}_3\text{P}_2$ compound. In practice, the Fermi-level should be pinned somewhere in between point A and B, depending on the Cd partial pressure.

On the experimental side, Selim and Kr{\"o}ger \cite{Selim77} found a sharp reduction in the hole concentration in samples annealed at 700$^\circ$ C under high cadmium partial pressures. These results are consistent with our findings which indicate the existence of a strong compensation between $(\text{P}_\text{Te})^{-1}$ and $(\text{P}_\text{Te}-\text{V}_\text{Te})^{+1}$ complexes under these conditions. In addition, the thermodynamic study performed by Yang et al. \cite{Yang15} suggests that the formation of AX centers (which previous theoretical results suggest that are responsible for most of the self-compensation) can be suppressed by the introduction of P at high temperature followed by rapid cooling. However, recent experimental observations \cite{Burst16} suggest the existence of a self-compensation mechanism independent of cooling rates.

According to our results, only a small increase in the V$_\text{oc}$ is expected with P doping alone.
Nevertheless, as the substitutional acceptor $(\text{P}_\text{Te})^{-1}$ is characterized by having a PHS at the VBM, it can be easily ionized at room temperature and could increase the hole density if the P concentration is high and the annealing is carried out under moderate Cd partial pressures. The structural data used in this study is provided as Supplementary Material [\onlinecite{Supp_material}].\\

\section{Summary}

In summary, we investigate the self-compensation mechanism in phosphorus-doped CdTe. Our study shows that the limiting factor for an  efficient \emph{p-}type doping is not due to the formation of AX centers. We find that, under Te-rich conditions, the limiting mechanism is due to self-compensation between P interstitials and substitutional $(\text{P}_\text{Te})^{-1}$ acceptors.
Morever, for Cd-rich growth conditions, we find that the self-compensation mechanism is due to the formation of $(\text{P}_\text{Te}-\text{V}_\text{Te})^{+1}$ complexes. Our results apply \emph{mutatis mutandis} to other group-V dopants in CdTe, suggesting an explanation for the \emph{p-}type doping limitations experimentally observed.

\vspace{0.5cm}
\begin{acknowledgments}
This work was supported by the Fondo Nacional de Investigaciones Cient\'ificas y Tecnol\'ogicas (FONDECYT, Chile) under grants No. 1170480 (W.O.) and 1171807 (E.M-P.). Powered@NLHPC: This research was partially supported by the supercomputing infrastructure of the NLHPC (ECM-02).
\end{acknowledgments}

\bibliographystyle{apsrev4-1}
\bibliography{ref}

\begin{thebibliography}{51}%
\makeatletter
\providecommand \@ifxundefined [1]{%
 \@ifx{#1\undefined}
}%
\providecommand \@ifnum [1]{%
 \ifnum #1\expandafter \@firstoftwo
 \else \expandafter \@secondoftwo
 \fi
}%
\providecommand \@ifx [1]{%
 \ifx #1\expandafter \@firstoftwo
 \else \expandafter \@secondoftwo
 \fi
}%
\providecommand \natexlab [1]{#1}%
\providecommand \enquote  [1]{``#1''}%
\providecommand \bibnamefont  [1]{#1}%
\providecommand \bibfnamefont [1]{#1}%
\providecommand \citenamefont [1]{#1}%
\providecommand \href@noop [0]{\@secondoftwo}%
\providecommand \href [0]{\begingroup \@sanitize@url \@href}%
\providecommand \@href[1]{\@@startlink{#1}\@@href}%
\providecommand \@@href[1]{\endgroup#1\@@endlink}%
\providecommand \@sanitize@url [0]{\catcode `\\12\catcode `\$12\catcode
  `\&12\catcode `\#12\catcode `\^12\catcode `\_12\catcode `\%12\relax}%
\providecommand \@@startlink[1]{}%
\providecommand \@@endlink[0]{}%
\providecommand \url  [0]{\begingroup\@sanitize@url \@url }%
\providecommand \@url [1]{\endgroup\@href {#1}{\urlprefix }}%
\providecommand \urlprefix  [0]{URL }%
\providecommand \Eprint [0]{\href }%
\providecommand \doibase [0]{http://dx.doi.org/}%
\providecommand \selectlanguage [0]{\@gobble}%
\providecommand \bibinfo  [0]{\@secondoftwo}%
\providecommand \bibfield  [0]{\@secondoftwo}%
\providecommand \translation [1]{[#1]}%
\providecommand \BibitemOpen [0]{}%
\providecommand \bibitemStop [0]{}%
\providecommand \bibitemNoStop [0]{.\EOS\space}%
\providecommand \EOS [0]{\spacefactor3000\relax}%
\providecommand \BibitemShut  [1]{\csname bibitem#1\endcsname}%
\let\auto@bib@innerbib\@empty
\bibitem [{\citenamefont {Li}\ \emph {et~al.}(2014)\citenamefont {Li},
  \citenamefont {Wu}, \citenamefont {Poplawsky}, \citenamefont {Pennycook},
  \citenamefont {Paudel}, \citenamefont {Yin}, \citenamefont {Haigh},
  \citenamefont {Oxley}, \citenamefont {Lupini}, \citenamefont {Al-Jassim}
  \emph {et~al.}}]{Chen14}%
  \BibitemOpen
  \bibfield  {author} {\bibinfo {author} {\bibfnamefont {C.}~\bibnamefont
  {Li}}, \bibinfo {author} {\bibfnamefont {Y.}~\bibnamefont {Wu}}, \bibinfo
  {author} {\bibfnamefont {J.}~\bibnamefont {Poplawsky}}, \bibinfo {author}
  {\bibfnamefont {T.~J.}\ \bibnamefont {Pennycook}}, \bibinfo {author}
  {\bibfnamefont {N.}~\bibnamefont {Paudel}}, \bibinfo {author} {\bibfnamefont
  {W.}~\bibnamefont {Yin}}, \bibinfo {author} {\bibfnamefont {S.~J.}\
  \bibnamefont {Haigh}}, \bibinfo {author} {\bibfnamefont {M.~P.}\ \bibnamefont
  {Oxley}}, \bibinfo {author} {\bibfnamefont {A.~R.}\ \bibnamefont {Lupini}},
  \bibinfo {author} {\bibfnamefont {M.}~\bibnamefont {Al-Jassim}},  \emph
  {et~al.},\ }\href {\doibase 10.1103/PhysRevLett.112.156103} {\bibfield
  {journal} {\bibinfo  {journal} {Phys. Rev. Lett.}\ }\textbf {\bibinfo
  {volume} {112}},\ \bibinfo {pages} {156103} (\bibinfo {year}
  {2014})}\BibitemShut {NoStop}%
\bibitem [{\citenamefont {Zhao}\ \emph {et~al.}(2016)\citenamefont {Zhao},
  \citenamefont {Boccard}, \citenamefont {Liu}, \citenamefont {Becker},
  \citenamefont {Zhao}, \citenamefont {Campbell}, \citenamefont {Suarez},
  \citenamefont {Lassise}, \citenamefont {Holman},\ and\ \citenamefont
  {Zhang}}]{Zhao16}%
  \BibitemOpen
  \bibfield  {author} {\bibinfo {author} {\bibfnamefont {Y.}~\bibnamefont
  {Zhao}}, \bibinfo {author} {\bibfnamefont {M.}~\bibnamefont {Boccard}},
  \bibinfo {author} {\bibfnamefont {S.}~\bibnamefont {Liu}}, \bibinfo {author}
  {\bibfnamefont {J.}~\bibnamefont {Becker}}, \bibinfo {author} {\bibfnamefont
  {X.-H.}\ \bibnamefont {Zhao}}, \bibinfo {author} {\bibfnamefont {C.~M.}\
  \bibnamefont {Campbell}}, \bibinfo {author} {\bibfnamefont {E.}~\bibnamefont
  {Suarez}}, \bibinfo {author} {\bibfnamefont {M.~B.}\ \bibnamefont {Lassise}},
  \bibinfo {author} {\bibfnamefont {Z.}~\bibnamefont {Holman}}, \ and\ \bibinfo
  {author} {\bibfnamefont {Y.-H.}\ \bibnamefont {Zhang}},\ }\href {\doibase
  10.1038/NENERGY.2016.67} {\bibfield  {journal} {\bibinfo  {journal} {Nat.
  Energy}\ }\textbf {\bibinfo {volume} {1}},\ \bibinfo {pages} {16067}
  (\bibinfo {year} {2016})}\BibitemShut {NoStop}%
\bibitem [{\citenamefont {Gessert}\ \emph {et~al.}(2013)\citenamefont
  {Gessert}, \citenamefont {Wei}, \citenamefont {Ma}, \citenamefont {Albin},
  \citenamefont {Dhere}, \citenamefont {Duenow}, \citenamefont {Kuciauskas},
  \citenamefont {Kanevce}, \citenamefont {Barnes}, \citenamefont {Burst} \emph
  {et~al.}}]{Gessert13}%
  \BibitemOpen
  \bibfield  {author} {\bibinfo {author} {\bibfnamefont {T.}~\bibnamefont
  {Gessert}}, \bibinfo {author} {\bibfnamefont {S.-H.}\ \bibnamefont {Wei}},
  \bibinfo {author} {\bibfnamefont {J.}~\bibnamefont {Ma}}, \bibinfo {author}
  {\bibfnamefont {D.}~\bibnamefont {Albin}}, \bibinfo {author} {\bibfnamefont
  {R.}~\bibnamefont {Dhere}}, \bibinfo {author} {\bibfnamefont
  {J.}~\bibnamefont {Duenow}}, \bibinfo {author} {\bibfnamefont
  {D.}~\bibnamefont {Kuciauskas}}, \bibinfo {author} {\bibfnamefont
  {A.}~\bibnamefont {Kanevce}}, \bibinfo {author} {\bibfnamefont
  {T.}~\bibnamefont {Barnes}}, \bibinfo {author} {\bibfnamefont
  {J.}~\bibnamefont {Burst}},  \emph {et~al.},\ }\href {\doibase
  10.1016/j.solmat.2013.05.055} {\bibfield  {journal} {\bibinfo  {journal}
  {Sol. Energ. Mat. Sol. Cells}\ }\textbf {\bibinfo {volume} {119}},\ \bibinfo
  {pages} {149 } (\bibinfo {year} {2013})}\BibitemShut {NoStop}%
\bibitem [{\citenamefont {Duenow}\ \emph {et~al.}(2016)\citenamefont {Duenow},
  \citenamefont {Burst}, \citenamefont {Albin}, \citenamefont {Reese},
  \citenamefont {Jensen}, \citenamefont {Johnston}, \citenamefont {Kuciauskas},
  \citenamefont {Swain}, \citenamefont {Ablekim}, \citenamefont {Lynn} \emph
  {et~al.}}]{Duenow16}%
  \BibitemOpen
  \bibfield  {author} {\bibinfo {author} {\bibfnamefont {J.~N.}\ \bibnamefont
  {Duenow}}, \bibinfo {author} {\bibfnamefont {J.~M.}\ \bibnamefont {Burst}},
  \bibinfo {author} {\bibfnamefont {D.~S.}\ \bibnamefont {Albin}}, \bibinfo
  {author} {\bibfnamefont {M.~O.}\ \bibnamefont {Reese}}, \bibinfo {author}
  {\bibfnamefont {S.~A.}\ \bibnamefont {Jensen}}, \bibinfo {author}
  {\bibfnamefont {S.~W.}\ \bibnamefont {Johnston}}, \bibinfo {author}
  {\bibfnamefont {D.}~\bibnamefont {Kuciauskas}}, \bibinfo {author}
  {\bibfnamefont {S.~K.}\ \bibnamefont {Swain}}, \bibinfo {author}
  {\bibfnamefont {T.}~\bibnamefont {Ablekim}}, \bibinfo {author} {\bibfnamefont
  {K.~G.}\ \bibnamefont {Lynn}},  \emph {et~al.},\ }\href {\doibase
  10.1109/JPHOTOV.2016.2598260} {\bibfield  {journal} {\bibinfo  {journal}
  {IEEE J. Photovolt.}\ }\textbf {\bibinfo {volume} {6}},\ \bibinfo {pages}
  {1641} (\bibinfo {year} {2016})}\BibitemShut {NoStop}%
\bibitem [{\citenamefont {Yang}\ \emph {et~al.}(2016)\citenamefont {Yang},
  \citenamefont {Yin}, \citenamefont {Park}, \citenamefont {Ma},\ and\
  \citenamefont {Wei}}]{Yang16}%
  \BibitemOpen
  \bibfield  {author} {\bibinfo {author} {\bibfnamefont {J.-H.}\ \bibnamefont
  {Yang}}, \bibinfo {author} {\bibfnamefont {W.-J.}\ \bibnamefont {Yin}},
  \bibinfo {author} {\bibfnamefont {J.-S.}\ \bibnamefont {Park}}, \bibinfo
  {author} {\bibfnamefont {J.}~\bibnamefont {Ma}}, \ and\ \bibinfo {author}
  {\bibfnamefont {S.-H.}\ \bibnamefont {Wei}},\ }\href {\doibase
  10.1088/0268-1242/31/8/083002} {\bibfield  {journal} {\bibinfo  {journal}
  {Semicond. Sci. Technol.}\ }\textbf {\bibinfo {volume} {31}},\ \bibinfo
  {pages} {083002} (\bibinfo {year} {2016})}\BibitemShut {NoStop}%
\bibitem [{\citenamefont {Flores}\ \emph
  {et~al.}(2016{\natexlab{a}})\citenamefont {Flores}, \citenamefont
  {Orellana},\ and\ \citenamefont {Men\'endez-Proupin}}]{Flores16_2}%
  \BibitemOpen
  \bibfield  {author} {\bibinfo {author} {\bibfnamefont {M.~A.}\ \bibnamefont
  {Flores}}, \bibinfo {author} {\bibfnamefont {W.}~\bibnamefont {Orellana}}, \
  and\ \bibinfo {author} {\bibfnamefont {E.}~\bibnamefont
  {Men\'endez-Proupin}},\ }\href {\doibase 10.1016/j.commatsci.2016.08.044}
  {\bibfield  {journal} {\bibinfo  {journal} {Comput. Mater. Sci.}\ }\textbf
  {\bibinfo {volume} {125}},\ \bibinfo {pages} {176 } (\bibinfo {year}
  {2016}{\natexlab{a}})}\BibitemShut {NoStop}%
\bibitem [{\citenamefont {Wickramaratne}\ \emph {et~al.}(2016)\citenamefont
  {Wickramaratne}, \citenamefont {Shen}, \citenamefont {Dreyer}, \citenamefont
  {Engel}, \citenamefont {Marsman}, \citenamefont {Kresse}, \citenamefont
  {Marcinkevi{\v{c}}ius}, \citenamefont {Alkauskas},\ and\ \citenamefont
  {Van~de Walle}}]{Wickramaratne16}%
  \BibitemOpen
  \bibfield  {author} {\bibinfo {author} {\bibfnamefont {D.}~\bibnamefont
  {Wickramaratne}}, \bibinfo {author} {\bibfnamefont {J.-X.}\ \bibnamefont
  {Shen}}, \bibinfo {author} {\bibfnamefont {C.~E.}\ \bibnamefont {Dreyer}},
  \bibinfo {author} {\bibfnamefont {M.}~\bibnamefont {Engel}}, \bibinfo
  {author} {\bibfnamefont {M.}~\bibnamefont {Marsman}}, \bibinfo {author}
  {\bibfnamefont {G.}~\bibnamefont {Kresse}}, \bibinfo {author} {\bibfnamefont
  {S.}~\bibnamefont {Marcinkevi{\v{c}}ius}}, \bibinfo {author} {\bibfnamefont
  {A.}~\bibnamefont {Alkauskas}}, \ and\ \bibinfo {author} {\bibfnamefont
  {C.~G.}\ \bibnamefont {Van~de Walle}},\ }\href {\doibase 10.1063/1.4964831}
  {\bibfield  {journal} {\bibinfo  {journal} {Appl. Phys. Lett.}\ }\textbf
  {\bibinfo {volume} {109}},\ \bibinfo {pages} {162107} (\bibinfo {year}
  {2016})}\BibitemShut {NoStop}%
\bibitem [{\citenamefont {Lyons}\ \emph {et~al.}(2014)\citenamefont {Lyons},
  \citenamefont {Janotti},\ and\ \citenamefont {Van~de Walle}}]{Lyons14}%
  \BibitemOpen
  \bibfield  {author} {\bibinfo {author} {\bibfnamefont {J.}~\bibnamefont
  {Lyons}}, \bibinfo {author} {\bibfnamefont {A.}~\bibnamefont {Janotti}}, \
  and\ \bibinfo {author} {\bibfnamefont {C.}~\bibnamefont {Van~de Walle}},\
  }\href {\doibase 10.1063/1.4838075} {\bibfield  {journal} {\bibinfo
  {journal} {J. Appl. Phys.}\ }\textbf {\bibinfo {volume} {115}},\ \bibinfo
  {pages} {012014} (\bibinfo {year} {2014})}\BibitemShut {NoStop}%
\bibitem [{\citenamefont {Kranz}\ \emph {et~al.}(2013)\citenamefont {Kranz},
  \citenamefont {Gretener}, \citenamefont {Perrenoud}, \citenamefont {Schmitt},
  \citenamefont {Pianezzi}, \citenamefont {La~Mattina}, \citenamefont
  {Bl{\"o}sch}, \citenamefont {Cheah}, \citenamefont {Chirila}, \citenamefont
  {Fella} \emph {et~al.}}]{Krantz13}%
  \BibitemOpen
  \bibfield  {author} {\bibinfo {author} {\bibfnamefont {L.}~\bibnamefont
  {Kranz}}, \bibinfo {author} {\bibfnamefont {C.}~\bibnamefont {Gretener}},
  \bibinfo {author} {\bibfnamefont {J.}~\bibnamefont {Perrenoud}}, \bibinfo
  {author} {\bibfnamefont {R.}~\bibnamefont {Schmitt}}, \bibinfo {author}
  {\bibfnamefont {F.}~\bibnamefont {Pianezzi}}, \bibinfo {author}
  {\bibfnamefont {F.}~\bibnamefont {La~Mattina}}, \bibinfo {author}
  {\bibfnamefont {P.}~\bibnamefont {Bl{\"o}sch}}, \bibinfo {author}
  {\bibfnamefont {E.}~\bibnamefont {Cheah}}, \bibinfo {author} {\bibfnamefont
  {A.}~\bibnamefont {Chirila}}, \bibinfo {author} {\bibfnamefont {C.~M.}\
  \bibnamefont {Fella}},  \emph {et~al.},\ }\href {\doibase 10.1038/ncomms3306}
  {\bibfield  {journal} {\bibinfo  {journal} {Nat. Commun.}\ }\textbf {\bibinfo
  {volume} {4}},\ \bibinfo {pages} {2306} (\bibinfo {year} {2013})}\BibitemShut
  {NoStop}%
\bibitem [{\citenamefont {Korcsm\'aros}\ \emph {et~al.}(2016)\citenamefont
  {Korcsm\'aros}, \citenamefont {Moravec}, \citenamefont {Grill}, \citenamefont
  {Musiienko},\ and\ \citenamefont {Ma\u{s}ek}}]{Korcsmaros16}%
  \BibitemOpen
  \bibfield  {author} {\bibinfo {author} {\bibfnamefont {G.}~\bibnamefont
  {Korcsm\'aros}}, \bibinfo {author} {\bibfnamefont {P.}~\bibnamefont
  {Moravec}}, \bibinfo {author} {\bibfnamefont {R.}~\bibnamefont {Grill}},
  \bibinfo {author} {\bibfnamefont {A.}~\bibnamefont {Musiienko}}, \ and\
  \bibinfo {author} {\bibfnamefont {K.}~\bibnamefont {Ma\u{s}ek}},\ }\href
  {\doibase 10.1016/j.jallcom.2016.03.233} {\bibfield  {journal} {\bibinfo
  {journal} {J. Alloy Compd.}\ }\textbf {\bibinfo {volume} {680}},\ \bibinfo
  {pages} {8 } (\bibinfo {year} {2016})}\BibitemShut {NoStop}%
\bibitem [{\citenamefont {Ablekim}\ \emph {et~al.}(2017)\citenamefont
  {Ablekim}, \citenamefont {Swain}, \citenamefont {Yin}, \citenamefont
  {Zaunbrecher}, \citenamefont {Burst}, \citenamefont {Barnes}, \citenamefont
  {Kuciauskas}, \citenamefont {Wei},\ and\ \citenamefont {Lynn}}]{Ablekim17}%
  \BibitemOpen
  \bibfield  {author} {\bibinfo {author} {\bibfnamefont {T.}~\bibnamefont
  {Ablekim}}, \bibinfo {author} {\bibfnamefont {S.~K.}\ \bibnamefont {Swain}},
  \bibinfo {author} {\bibfnamefont {W.-J.}\ \bibnamefont {Yin}}, \bibinfo
  {author} {\bibfnamefont {K.}~\bibnamefont {Zaunbrecher}}, \bibinfo {author}
  {\bibfnamefont {J.}~\bibnamefont {Burst}}, \bibinfo {author} {\bibfnamefont
  {T.~M.}\ \bibnamefont {Barnes}}, \bibinfo {author} {\bibfnamefont
  {D.}~\bibnamefont {Kuciauskas}}, \bibinfo {author} {\bibfnamefont {S.-H.}\
  \bibnamefont {Wei}}, \ and\ \bibinfo {author} {\bibfnamefont {K.~G.}\
  \bibnamefont {Lynn}},\ }\href {\doibase 10.1038/s41598-017-04719-0}
  {\bibfield  {journal} {\bibinfo  {journal} {Sci. Rep.}\ }\textbf {\bibinfo
  {volume} {7}},\ \bibinfo {pages} {4563} (\bibinfo {year} {2017})}\BibitemShut
  {NoStop}%
\bibitem [{\citenamefont {Mandel}(1964)}]{Mandel64}%
  \BibitemOpen
  \bibfield  {author} {\bibinfo {author} {\bibfnamefont {G.}~\bibnamefont
  {Mandel}},\ }\href {\doibase 10.1103/PhysRev.134.A1073} {\bibfield  {journal}
  {\bibinfo  {journal} {Phys. Rev.}\ }\textbf {\bibinfo {volume} {134}},\
  \bibinfo {pages} {A1073} (\bibinfo {year} {1964})}\BibitemShut {NoStop}%
\bibitem [{\citenamefont {Krasikov}\ \emph {et~al.}(2013)\citenamefont
  {Krasikov}, \citenamefont {Knizhnik}, \citenamefont {Potapkin},\ and\
  \citenamefont {Sommerer}}]{Krasikov13}%
  \BibitemOpen
  \bibfield  {author} {\bibinfo {author} {\bibfnamefont {D.}~\bibnamefont
  {Krasikov}}, \bibinfo {author} {\bibfnamefont {A.}~\bibnamefont {Knizhnik}},
  \bibinfo {author} {\bibfnamefont {B.}~\bibnamefont {Potapkin}}, \ and\
  \bibinfo {author} {\bibfnamefont {T.}~\bibnamefont {Sommerer}},\ }\href
  {\doibase 10.1088/0268-1242/28/12/125019} {\bibfield  {journal} {\bibinfo
  {journal} {Semicond. Sci. Technol.}\ }\textbf {\bibinfo {volume} {28}},\
  \bibinfo {pages} {125019} (\bibinfo {year} {2013})}\BibitemShut {NoStop}%
\bibitem [{\citenamefont {Desnica}(1998)}]{Desnica98}%
  \BibitemOpen
  \bibfield  {author} {\bibinfo {author} {\bibfnamefont {U.}~\bibnamefont
  {Desnica}},\ }\href {\doibase 10.1016/S0960-8974(98)00011-4} {\bibfield
  {journal} {\bibinfo  {journal} {Prog. Cryst. Growth Charact. Mater.}\
  }\textbf {\bibinfo {volume} {36}},\ \bibinfo {pages} {291 } (\bibinfo {year}
  {1998})}\BibitemShut {NoStop}%
\bibitem [{\citenamefont {Wei}\ and\ \citenamefont {Zhang}(2002)}]{Wei02}%
  \BibitemOpen
  \bibfield  {author} {\bibinfo {author} {\bibfnamefont {S.-H.}\ \bibnamefont
  {Wei}}\ and\ \bibinfo {author} {\bibfnamefont {S.~B.}\ \bibnamefont
  {Zhang}},\ }\href {\doibase 10.1103/PhysRevB.66.155211} {\bibfield  {journal}
  {\bibinfo  {journal} {Phys. Rev. B}\ }\textbf {\bibinfo {volume} {66}},\
  \bibinfo {pages} {155211} (\bibinfo {year} {2002})}\BibitemShut {NoStop}%
\bibitem [{\citenamefont {Jones}\ \emph {et~al.}(1992)\citenamefont {Jones},
  \citenamefont {Stewart},\ and\ \citenamefont {Mullin}}]{Jones92}%
  \BibitemOpen
  \bibfield  {author} {\bibinfo {author} {\bibfnamefont {E.}~\bibnamefont
  {Jones}}, \bibinfo {author} {\bibfnamefont {N.}~\bibnamefont {Stewart}}, \
  and\ \bibinfo {author} {\bibfnamefont {J.}~\bibnamefont {Mullin}},\ }\href
  {\doibase 10.1016/0022-0248(92)90753-6} {\bibfield  {journal} {\bibinfo
  {journal} {J. Cryst. Growth}\ }\textbf {\bibinfo {volume} {117}},\ \bibinfo
  {pages} {244 } (\bibinfo {year} {1992})}\BibitemShut {NoStop}%
\bibitem [{\citenamefont {Grecu}\ \emph {et~al.}(2000)\citenamefont {Grecu},
  \citenamefont {Compaan}, \citenamefont {Young}, \citenamefont {Jayamaha},\
  and\ \citenamefont {Rose}}]{Grecua00}%
  \BibitemOpen
  \bibfield  {author} {\bibinfo {author} {\bibfnamefont {D.}~\bibnamefont
  {Grecu}}, \bibinfo {author} {\bibfnamefont {A.}~\bibnamefont {Compaan}},
  \bibinfo {author} {\bibfnamefont {D.}~\bibnamefont {Young}}, \bibinfo
  {author} {\bibfnamefont {U.}~\bibnamefont {Jayamaha}}, \ and\ \bibinfo
  {author} {\bibfnamefont {D.}~\bibnamefont {Rose}},\ }\href {\doibase
  10.1063/1.1287414} {\bibfield  {journal} {\bibinfo  {journal} {J. Appl.
  Phys.}\ }\textbf {\bibinfo {volume} {88}},\ \bibinfo {pages} {2490} (\bibinfo
  {year} {2000})}\BibitemShut {NoStop}%
\bibitem [{\citenamefont {Dzhafarov}\ \emph {et~al.}(2005)\citenamefont
  {Dzhafarov}, \citenamefont {Yesilkaya}, \citenamefont {Canli},\ and\
  \citenamefont {Caliskan}}]{Dzhafarov05}%
  \BibitemOpen
  \bibfield  {author} {\bibinfo {author} {\bibfnamefont {T.}~\bibnamefont
  {Dzhafarov}}, \bibinfo {author} {\bibfnamefont {S.}~\bibnamefont
  {Yesilkaya}}, \bibinfo {author} {\bibfnamefont {N.~Y.}\ \bibnamefont
  {Canli}}, \ and\ \bibinfo {author} {\bibfnamefont {M.}~\bibnamefont
  {Caliskan}},\ }\href {\doibase 10.1016/j.solmat.2004.05.007} {\bibfield
  {journal} {\bibinfo  {journal} {Sol. Energ. Mat. Sol. Cells}\ }\textbf
  {\bibinfo {volume} {85}},\ \bibinfo {pages} {371} (\bibinfo {year}
  {2005})}\BibitemShut {NoStop}%
\bibitem [{\citenamefont {Yang}\ \emph {et~al.}(2017)\citenamefont {Yang},
  \citenamefont {Metzger},\ and\ \citenamefont {Wei}}]{Yang17}%
  \BibitemOpen
  \bibfield  {author} {\bibinfo {author} {\bibfnamefont {J.-H.}\ \bibnamefont
  {Yang}}, \bibinfo {author} {\bibfnamefont {W.~K.}\ \bibnamefont {Metzger}}, \
  and\ \bibinfo {author} {\bibfnamefont {S.-H.}\ \bibnamefont {Wei}},\ }\href
  {\doibase 10.1063/1.4986077} {\bibfield  {journal} {\bibinfo  {journal}
  {Appl. Phys. Lett.}\ }\textbf {\bibinfo {volume} {111}},\ \bibinfo {pages}
  {042106} (\bibinfo {year} {2017})}\BibitemShut {NoStop}%
\bibitem [{\citenamefont {Burst}\ \emph {et~al.}(2016)\citenamefont {Burst},
  \citenamefont {Duenow}, \citenamefont {Albin}, \citenamefont {Colegrove},
  \citenamefont {Reese}, \citenamefont {Aguiar}, \citenamefont {Jiang},
  \citenamefont {Patel}, \citenamefont {Al-Jassim}, \citenamefont {Kuciauskas}
  \emph {et~al.}}]{Burst16}%
  \BibitemOpen
  \bibfield  {author} {\bibinfo {author} {\bibfnamefont {J.~M.}\ \bibnamefont
  {Burst}}, \bibinfo {author} {\bibfnamefont {J.~N.}\ \bibnamefont {Duenow}},
  \bibinfo {author} {\bibfnamefont {D.~S.}\ \bibnamefont {Albin}}, \bibinfo
  {author} {\bibfnamefont {E.}~\bibnamefont {Colegrove}}, \bibinfo {author}
  {\bibfnamefont {M.~O.}\ \bibnamefont {Reese}}, \bibinfo {author}
  {\bibfnamefont {J.~A.}\ \bibnamefont {Aguiar}}, \bibinfo {author}
  {\bibfnamefont {C.-S.}\ \bibnamefont {Jiang}}, \bibinfo {author}
  {\bibfnamefont {M.}~\bibnamefont {Patel}}, \bibinfo {author} {\bibfnamefont
  {M.~M.}\ \bibnamefont {Al-Jassim}}, \bibinfo {author} {\bibfnamefont
  {D.}~\bibnamefont {Kuciauskas}},  \emph {et~al.},\ }\href {\doibase
  10.1038/nenergy.2016.15} {\bibfield  {journal} {\bibinfo  {journal} {Nat.
  Energy.}\ }\textbf {\bibinfo {volume} {1}},\ \bibinfo {pages} {16015}
  (\bibinfo {year} {2016})}\BibitemShut {NoStop}%
\bibitem [{\citenamefont {Nagaoka}\ \emph {et~al.}(2017)\citenamefont
  {Nagaoka}, \citenamefont {Han}, \citenamefont {Misra}, \citenamefont
  {Wilenski}, \citenamefont {Sparks},\ and\ \citenamefont
  {Scarpulla}}]{Nagaoka17}%
  \BibitemOpen
  \bibfield  {author} {\bibinfo {author} {\bibfnamefont {A.}~\bibnamefont
  {Nagaoka}}, \bibinfo {author} {\bibfnamefont {K.-B.}\ \bibnamefont {Han}},
  \bibinfo {author} {\bibfnamefont {S.}~\bibnamefont {Misra}}, \bibinfo
  {author} {\bibfnamefont {T.}~\bibnamefont {Wilenski}}, \bibinfo {author}
  {\bibfnamefont {T.~D.}\ \bibnamefont {Sparks}}, \ and\ \bibinfo {author}
  {\bibfnamefont {M.~A.}\ \bibnamefont {Scarpulla}},\ }\href {\doibase
  10.1016/j.jcrysgro.2017.03.002} {\bibfield  {journal} {\bibinfo  {journal}
  {J. Crys. Growth}\ }\textbf {\bibinfo {volume} {467}},\ \bibinfo {pages} {6}
  (\bibinfo {year} {2017})}\BibitemShut {NoStop}%
\bibitem [{\citenamefont {Yang}\ \emph {et~al.}(2015)\citenamefont {Yang},
  \citenamefont {Yin}, \citenamefont {Park}, \citenamefont {Burst},
  \citenamefont {Metzger}, \citenamefont {Gessert}, \citenamefont {Barnes},\
  and\ \citenamefont {Wei}}]{Yang15}%
  \BibitemOpen
  \bibfield  {author} {\bibinfo {author} {\bibfnamefont {J.-H.}\ \bibnamefont
  {Yang}}, \bibinfo {author} {\bibfnamefont {W.-J.}\ \bibnamefont {Yin}},
  \bibinfo {author} {\bibfnamefont {J.-S.}\ \bibnamefont {Park}}, \bibinfo
  {author} {\bibfnamefont {J.}~\bibnamefont {Burst}}, \bibinfo {author}
  {\bibfnamefont {W.~K.}\ \bibnamefont {Metzger}}, \bibinfo {author}
  {\bibfnamefont {T.}~\bibnamefont {Gessert}}, \bibinfo {author} {\bibfnamefont
  {T.}~\bibnamefont {Barnes}}, \ and\ \bibinfo {author} {\bibfnamefont {S.-H.}\
  \bibnamefont {Wei}},\ }\href {\doibase 10.1063/1.4926748} {\bibfield
  {journal} {\bibinfo  {journal} {J. Appl. Phys.}\ }\textbf {\bibinfo {volume}
  {118}},\ \bibinfo {pages} {025102} (\bibinfo {year} {2015})}\BibitemShut
  {NoStop}%
\bibitem [{\citenamefont {Colegrove}\ \emph {et~al.}(2016)\citenamefont
  {Colegrove}, \citenamefont {Harvey}, \citenamefont {Yang}, \citenamefont
  {Burst}, \citenamefont {Albin}, \citenamefont {Wei},\ and\ \citenamefont
  {Metzger}}]{Colegrove16}%
  \BibitemOpen
  \bibfield  {author} {\bibinfo {author} {\bibfnamefont {E.}~\bibnamefont
  {Colegrove}}, \bibinfo {author} {\bibfnamefont {S.~P.}\ \bibnamefont
  {Harvey}}, \bibinfo {author} {\bibfnamefont {J.-H.}\ \bibnamefont {Yang}},
  \bibinfo {author} {\bibfnamefont {J.~M.}\ \bibnamefont {Burst}}, \bibinfo
  {author} {\bibfnamefont {D.~S.}\ \bibnamefont {Albin}}, \bibinfo {author}
  {\bibfnamefont {S.-H.}\ \bibnamefont {Wei}}, \ and\ \bibinfo {author}
  {\bibfnamefont {W.~K.}\ \bibnamefont {Metzger}},\ }\href {\doibase
  10.1103/PhysRevApplied.5.054014} {\bibfield  {journal} {\bibinfo  {journal}
  {Phys. Rev. Applied}\ }\textbf {\bibinfo {volume} {5}},\ \bibinfo {pages}
  {054014} (\bibinfo {year} {2016})}\BibitemShut {NoStop}%
\bibitem [{\citenamefont {Colegrove}\ \emph {et~al.}(2017)\citenamefont
  {Colegrove}, \citenamefont {Harvey}, \citenamefont {Yang}, \citenamefont
  {Burst}, \citenamefont {Duenow}, \citenamefont {Albin}, \citenamefont {Wei},\
  and\ \citenamefont {Metzger}}]{Colegrove17}%
  \BibitemOpen
  \bibfield  {author} {\bibinfo {author} {\bibfnamefont {E.}~\bibnamefont
  {Colegrove}}, \bibinfo {author} {\bibfnamefont {S.~P.}\ \bibnamefont
  {Harvey}}, \bibinfo {author} {\bibfnamefont {J.~H.}\ \bibnamefont {Yang}},
  \bibinfo {author} {\bibfnamefont {J.~M.}\ \bibnamefont {Burst}}, \bibinfo
  {author} {\bibfnamefont {J.~N.}\ \bibnamefont {Duenow}}, \bibinfo {author}
  {\bibfnamefont {D.~S.}\ \bibnamefont {Albin}}, \bibinfo {author}
  {\bibfnamefont {S.~H.}\ \bibnamefont {Wei}}, \ and\ \bibinfo {author}
  {\bibfnamefont {W.~K.}\ \bibnamefont {Metzger}},\ }\href {\doibase
  10.1109/JPHOTOV.2017.2655033} {\bibfield  {journal} {\bibinfo  {journal}
  {IEEE J. Photovolt.}\ }\textbf {\bibinfo {volume} {7}},\ \bibinfo {pages}
  {870} (\bibinfo {year} {2017})}\BibitemShut {NoStop}%
\bibitem [{\citenamefont {Perdew}\ \emph {et~al.}(1996)\citenamefont {Perdew},
  \citenamefont {Burke},\ and\ \citenamefont {Ernzerhof}}]{Perdew96}%
  \BibitemOpen
  \bibfield  {author} {\bibinfo {author} {\bibfnamefont {J.~P.}\ \bibnamefont
  {Perdew}}, \bibinfo {author} {\bibfnamefont {K.}~\bibnamefont {Burke}}, \
  and\ \bibinfo {author} {\bibfnamefont {M.}~\bibnamefont {Ernzerhof}},\ }\href
  {\doibase 10.1103/PhysRevLett.77.3865} {\bibfield  {journal} {\bibinfo
  {journal} {Phys. Rev. Lett.}\ }\textbf {\bibinfo {volume} {77}},\ \bibinfo
  {pages} {3865} (\bibinfo {year} {1996})}\BibitemShut {NoStop}%
\bibitem [{\citenamefont {Giannozzi}\ \emph {et~al.}(2009)\citenamefont
  {Giannozzi}, \citenamefont {Baroni}, \citenamefont {Bonini}, \citenamefont
  {Calandra}, \citenamefont {Car}, \citenamefont {Cavazzoni}, \citenamefont
  {Ceresoli}, \citenamefont {Chiarotti}, \citenamefont {Cococcioni},
  \citenamefont {Dabo} \emph {et~al.}}]{Giannozzi2009}%
  \BibitemOpen
  \bibfield  {author} {\bibinfo {author} {\bibfnamefont {P.}~\bibnamefont
  {Giannozzi}}, \bibinfo {author} {\bibfnamefont {S.}~\bibnamefont {Baroni}},
  \bibinfo {author} {\bibfnamefont {N.}~\bibnamefont {Bonini}}, \bibinfo
  {author} {\bibfnamefont {M.}~\bibnamefont {Calandra}}, \bibinfo {author}
  {\bibfnamefont {R.}~\bibnamefont {Car}}, \bibinfo {author} {\bibfnamefont
  {C.}~\bibnamefont {Cavazzoni}}, \bibinfo {author} {\bibfnamefont
  {D.}~\bibnamefont {Ceresoli}}, \bibinfo {author} {\bibfnamefont {G.~L.}\
  \bibnamefont {Chiarotti}}, \bibinfo {author} {\bibfnamefont {M.}~\bibnamefont
  {Cococcioni}}, \bibinfo {author} {\bibfnamefont {I.}~\bibnamefont {Dabo}},
  \emph {et~al.},\ }\href {http://dx.doi.org/10.1088/0953-8984/21/39/395502}
  {\bibfield  {journal} {\bibinfo  {journal} {J. Phys.: Condens. Matter}\
  }\textbf {\bibinfo {volume} {21}},\ \bibinfo {pages} {395502} (\bibinfo
  {year} {2009})}\BibitemShut {NoStop}%
\bibitem [{\citenamefont {{K. Garrity, J. Bennett, K. Rabe, and D.
  Vanderbilt}}(2014)}]{Garrity2014}%
  \BibitemOpen
  \bibfield  {author} {\bibinfo {author} {\bibnamefont {{K. Garrity, J.
  Bennett, K. Rabe, and D. Vanderbilt}}},\ }\href {\doibase
  10.1088/0953-8984/21/39/395502} {\bibfield  {journal} {\bibinfo  {journal}
  {Comput. Mater. Sci.}\ }\textbf {\bibinfo {volume} {81}},\ \bibinfo {pages}
  {446} (\bibinfo {year} {2014})}\BibitemShut {NoStop}%
\bibitem [{\citenamefont {Lany}\ and\ \citenamefont {Zunger}(2008)}]{Lany08}%
  \BibitemOpen
  \bibfield  {author} {\bibinfo {author} {\bibfnamefont {S.}~\bibnamefont
  {Lany}}\ and\ \bibinfo {author} {\bibfnamefont {A.}~\bibnamefont {Zunger}},\
  }\href {\doibase 10.1103/PhysRevB.78.235104} {\bibfield  {journal} {\bibinfo
  {journal} {Phys. Rev. B}\ }\textbf {\bibinfo {volume} {78}},\ \bibinfo
  {pages} {235104} (\bibinfo {year} {2008})}\BibitemShut {NoStop}%
\bibitem [{\citenamefont {Kioupakis}\ \emph {et~al.}(2008)\citenamefont
  {Kioupakis}, \citenamefont {Zhang}, \citenamefont {Cohen},\ and\
  \citenamefont {Louie}}]{Kioupakis08}%
  \BibitemOpen
  \bibfield  {author} {\bibinfo {author} {\bibfnamefont {E.}~\bibnamefont
  {Kioupakis}}, \bibinfo {author} {\bibfnamefont {P.}~\bibnamefont {Zhang}},
  \bibinfo {author} {\bibfnamefont {M.~L.}\ \bibnamefont {Cohen}}, \ and\
  \bibinfo {author} {\bibfnamefont {S.~G.}\ \bibnamefont {Louie}},\ }\href
  {\doibase 10.1103/PhysRevB.77.155114} {\bibfield  {journal} {\bibinfo
  {journal} {Phys. Rev. B}\ }\textbf {\bibinfo {volume} {77}},\ \bibinfo
  {pages} {155114} (\bibinfo {year} {2008})}\BibitemShut {NoStop}%
\bibitem [{\citenamefont {Gonze}\ \emph {et~al.}(2009)\citenamefont {Gonze},
  \citenamefont {Amadon}, \citenamefont {Anglade}, \citenamefont {Beuken},
  \citenamefont {Bottin}, \citenamefont {Boulanger}, \citenamefont {Bruneval},
  \citenamefont {Caliste}, \citenamefont {Caracas}, \citenamefont
  {C{\^o}t{\'e}} \emph {et~al.}}]{Gonze2009}%
  \BibitemOpen
  \bibfield  {author} {\bibinfo {author} {\bibfnamefont {X.}~\bibnamefont
  {Gonze}}, \bibinfo {author} {\bibfnamefont {B.}~\bibnamefont {Amadon}},
  \bibinfo {author} {\bibfnamefont {P.-M.}\ \bibnamefont {Anglade}}, \bibinfo
  {author} {\bibfnamefont {J.-M.}\ \bibnamefont {Beuken}}, \bibinfo {author}
  {\bibfnamefont {F.}~\bibnamefont {Bottin}}, \bibinfo {author} {\bibfnamefont
  {P.}~\bibnamefont {Boulanger}}, \bibinfo {author} {\bibfnamefont
  {F.}~\bibnamefont {Bruneval}}, \bibinfo {author} {\bibfnamefont
  {D.}~\bibnamefont {Caliste}}, \bibinfo {author} {\bibfnamefont
  {R.}~\bibnamefont {Caracas}}, \bibinfo {author} {\bibfnamefont
  {M.}~\bibnamefont {C{\^o}t{\'e}}},  \emph {et~al.},\ }\href {\doibase
  10.1016/j.cpc.2009.07.007} {\bibfield  {journal} {\bibinfo  {journal}
  {Comput. Phys. Commun.}\ }\textbf {\bibinfo {volume} {180}},\ \bibinfo
  {pages} {2582 } (\bibinfo {year} {2009})}\BibitemShut {NoStop}%
\bibitem [{\citenamefont {Gonze}\ \emph {et~al.}(2016)\citenamefont {Gonze},
  \citenamefont {Jollet}, \citenamefont {Araujo}, \citenamefont {Adams},
  \citenamefont {Amadon}, \citenamefont {Applencourt}, \citenamefont {Audouze},
  \citenamefont {Beuken}, \citenamefont {Bieder}, \citenamefont {Bokhanchuk}
  \emph {et~al.}}]{Gonze16}%
  \BibitemOpen
  \bibfield  {author} {\bibinfo {author} {\bibfnamefont {X.}~\bibnamefont
  {Gonze}}, \bibinfo {author} {\bibfnamefont {F.}~\bibnamefont {Jollet}},
  \bibinfo {author} {\bibfnamefont {F.~A.}\ \bibnamefont {Araujo}}, \bibinfo
  {author} {\bibfnamefont {D.}~\bibnamefont {Adams}}, \bibinfo {author}
  {\bibfnamefont {B.}~\bibnamefont {Amadon}}, \bibinfo {author} {\bibfnamefont
  {T.}~\bibnamefont {Applencourt}}, \bibinfo {author} {\bibfnamefont
  {C.}~\bibnamefont {Audouze}}, \bibinfo {author} {\bibfnamefont {J.-M.}\
  \bibnamefont {Beuken}}, \bibinfo {author} {\bibfnamefont {J.}~\bibnamefont
  {Bieder}}, \bibinfo {author} {\bibfnamefont {A.}~\bibnamefont {Bokhanchuk}},
  \emph {et~al.},\ }\href {\doibase 10.1016/j.cpc.2016.04.003} {\bibfield
  {journal} {\bibinfo  {journal} {Comput. Phys. Commun.}\ }\textbf {\bibinfo
  {volume} {205}},\ \bibinfo {pages} {106 } (\bibinfo {year}
  {2016})}\BibitemShut {NoStop}%
\bibitem [{\citenamefont {Bruneval}\ and\ \citenamefont
  {Gonze}(2008)}]{Bruneval08}%
  \BibitemOpen
  \bibfield  {author} {\bibinfo {author} {\bibfnamefont {F.}~\bibnamefont
  {Bruneval}}\ and\ \bibinfo {author} {\bibfnamefont {X.}~\bibnamefont
  {Gonze}},\ }\href {\doibase 10.1103/PhysRevB.78.085125} {\bibfield  {journal}
  {\bibinfo  {journal} {Phys. Rev. B}\ }\textbf {\bibinfo {volume} {78}},\
  \bibinfo {pages} {085125} (\bibinfo {year} {2008})}\BibitemShut {NoStop}%
\bibitem [{\citenamefont {Godby}\ and\ \citenamefont {Needs}(1989)}]{Godby89}%
  \BibitemOpen
  \bibfield  {author} {\bibinfo {author} {\bibfnamefont {R.}~\bibnamefont
  {Godby}}\ and\ \bibinfo {author} {\bibfnamefont {R.}~\bibnamefont {Needs}},\
  }\href {\doibase 10.1103/PhysRevLett.62.1169} {\bibfield  {journal} {\bibinfo
   {journal} {hys. Rev. Lett}\ }\textbf {\bibinfo {volume} {62}},\ \bibinfo
  {pages} {1169} (\bibinfo {year} {1989})}\BibitemShut {NoStop}%
\bibitem [{\citenamefont {Freysoldt}\ \emph {et~al.}(2014)\citenamefont
  {Freysoldt}, \citenamefont {Grabowski}, \citenamefont {Hickel}, \citenamefont
  {Neugebauer}, \citenamefont {Kresse}, \citenamefont {Janotti},\ and\
  \citenamefont {Van~de Walle}}]{Freysoldt14}%
  \BibitemOpen
  \bibfield  {author} {\bibinfo {author} {\bibfnamefont {C.}~\bibnamefont
  {Freysoldt}}, \bibinfo {author} {\bibfnamefont {B.}~\bibnamefont
  {Grabowski}}, \bibinfo {author} {\bibfnamefont {T.}~\bibnamefont {Hickel}},
  \bibinfo {author} {\bibfnamefont {J.}~\bibnamefont {Neugebauer}}, \bibinfo
  {author} {\bibfnamefont {G.}~\bibnamefont {Kresse}}, \bibinfo {author}
  {\bibfnamefont {A.}~\bibnamefont {Janotti}}, \ and\ \bibinfo {author}
  {\bibfnamefont {C.~G.}\ \bibnamefont {Van~de Walle}},\ }\href {\doibase
  10.1103/RevModPhys.86.253} {\bibfield  {journal} {\bibinfo  {journal} {Rev.
  Mod. Phys.}\ }\textbf {\bibinfo {volume} {86}},\ \bibinfo {pages} {253}
  (\bibinfo {year} {2014})}\BibitemShut {NoStop}%
\bibitem [{\citenamefont {Jain}\ \emph {et~al.}(2011)\citenamefont {Jain},
  \citenamefont {Chelikowsky},\ and\ \citenamefont {Louie}}]{Jain11}%
  \BibitemOpen
  \bibfield  {author} {\bibinfo {author} {\bibfnamefont {M.}~\bibnamefont
  {Jain}}, \bibinfo {author} {\bibfnamefont {J.~R.}\ \bibnamefont
  {Chelikowsky}}, \ and\ \bibinfo {author} {\bibfnamefont {S.~G.}\ \bibnamefont
  {Louie}},\ }\href {\doibase 10.1103/PhysRevLett.107.216803} {\bibfield
  {journal} {\bibinfo  {journal} {Phys. Rev. Lett.}\ }\textbf {\bibinfo
  {volume} {107}},\ \bibinfo {pages} {216803} (\bibinfo {year}
  {2011})}\BibitemShut {NoStop}%
\bibitem [{\citenamefont {Flores}\ \emph
  {et~al.}(2016{\natexlab{b}})\citenamefont {Flores}, \citenamefont
  {Orellana},\ and\ \citenamefont {Men\'endez-Proupin}}]{Flores16_1}%
  \BibitemOpen
  \bibfield  {author} {\bibinfo {author} {\bibfnamefont {M.~A.}\ \bibnamefont
  {Flores}}, \bibinfo {author} {\bibfnamefont {W.}~\bibnamefont {Orellana}}, \
  and\ \bibinfo {author} {\bibfnamefont {E.}~\bibnamefont
  {Men\'endez-Proupin}},\ }\href {\doibase 10.1103/PhysRevB.93.184103}
  {\bibfield  {journal} {\bibinfo  {journal} {Phys. Rev. B}\ }\textbf {\bibinfo
  {volume} {93}},\ \bibinfo {pages} {184103} (\bibinfo {year}
  {2016}{\natexlab{b}})}\BibitemShut {NoStop}%
\bibitem [{\citenamefont {Flores}(2017)}]{Flores17_2}%
  \BibitemOpen
  \bibfield  {author} {\bibinfo {author} {\bibfnamefont {M.~A.}\ \bibnamefont
  {Flores}},\ }\href@noop {} {\enquote {\bibinfo {title} {\text{Defect}
  properties of \text{Sn- and Ge-} doped \text{ZnTe}: suitability for
  intermediate-band solar cells},}\ } (\bibinfo {year} {2017}),\ \Eprint
  {http://arxiv.org/abs/arXiv:1709.03277} {arXiv:1709.03277} \BibitemShut
  {NoStop}%
\bibitem [{\citenamefont {Zhang}(2002)}]{Zhang02}%
  \BibitemOpen
  \bibfield  {author} {\bibinfo {author} {\bibfnamefont {S.~B.}\ \bibnamefont
  {Zhang}},\ }\href {http://stacks.iop.org/0953-8984/14/i=34/a=201} {\bibfield
  {journal} {\bibinfo  {journal} {J. Phys. Condens. Matter}\ }\textbf {\bibinfo
  {volume} {14}},\ \bibinfo {pages} {R881} (\bibinfo {year}
  {2002})}\BibitemShut {NoStop}%
\bibitem [{\citenamefont {Lany}\ \emph {et~al.}(2001)\citenamefont {Lany},
  \citenamefont {Ostheimer}, \citenamefont {Wolf},\ and\ \citenamefont
  {Wichert}}]{Lany01}%
  \BibitemOpen
  \bibfield  {author} {\bibinfo {author} {\bibfnamefont {S.}~\bibnamefont
  {Lany}}, \bibinfo {author} {\bibfnamefont {V.}~\bibnamefont {Ostheimer}},
  \bibinfo {author} {\bibfnamefont {H.}~\bibnamefont {Wolf}}, \ and\ \bibinfo
  {author} {\bibfnamefont {T.}~\bibnamefont {Wichert}},\ }\href {\doibase
  10.1016/S0921-4526(01)00841-9} {\bibfield  {journal} {\bibinfo  {journal}
  {Physica B}\ }\textbf {\bibinfo {volume} {308}},\ \bibinfo {pages} {958 }
  (\bibinfo {year} {2001})}\BibitemShut {NoStop}%
\bibitem [{\citenamefont {Lany}\ and\ \citenamefont {Zunger}(2010)}]{Lany10}%
  \BibitemOpen
  \bibfield  {author} {\bibinfo {author} {\bibfnamefont {S.}~\bibnamefont
  {Lany}}\ and\ \bibinfo {author} {\bibfnamefont {A.}~\bibnamefont {Zunger}},\
  }\href {\doibase 10.1063/1.3383236} {\bibfield  {journal} {\bibinfo
  {journal} {Appl. Phys. Lett.}\ }\textbf {\bibinfo {volume} {96}},\ \bibinfo
  {pages} {142114} (\bibinfo {year} {2010})}\BibitemShut {NoStop}%
\bibitem [{\citenamefont {Bl\"ochl}(1994)}]{Bloch94}%
  \BibitemOpen
  \bibfield  {author} {\bibinfo {author} {\bibfnamefont {P.~E.}\ \bibnamefont
  {Bl\"ochl}},\ }\href {\doibase 10.1103/PhysRevB.50.17953} {\bibfield
  {journal} {\bibinfo  {journal} {Phys. Rev. B}\ }\textbf {\bibinfo {volume}
  {50}},\ \bibinfo {pages} {17953} (\bibinfo {year} {1994})}\BibitemShut
  {NoStop}%
\bibitem [{\citenamefont {{A. Dal Corso}}(2014)}]{DalCorso14}%
  \BibitemOpen
  \bibfield  {author} {\bibinfo {author} {\bibnamefont {{A. Dal Corso}}},\
  }\href {\doibase 10.1016/j.commatsci.2014.07.043} {\bibfield  {journal}
  {\bibinfo  {journal} {Comput. Mater. Sci.}\ }\textbf {\bibinfo {volume}
  {95}},\ \bibinfo {pages} {337 } (\bibinfo {year} {2014})}\BibitemShut
  {NoStop}%
\bibitem [{\citenamefont {Heyd}\ \emph {et~al.}(2006)\citenamefont {Heyd},
  \citenamefont {Scuseria},\ and\ \citenamefont {Ernzerhof}}]{Heyd06}%
  \BibitemOpen
  \bibfield  {author} {\bibinfo {author} {\bibfnamefont {J.}~\bibnamefont
  {Heyd}}, \bibinfo {author} {\bibfnamefont {G.}~\bibnamefont {Scuseria}}, \
  and\ \bibinfo {author} {\bibfnamefont {M.}~\bibnamefont {Ernzerhof}},\ }\href
  {\doibase 10.1063/1.2204597} {\bibfield  {journal} {\bibinfo  {journal} {J.
  Chem. Phys.}\ }\textbf {\bibinfo {volume} {124}},\ \bibinfo {pages} {219906}
  (\bibinfo {year} {2006})}\BibitemShut {NoStop}%
\bibitem [{\citenamefont {Broqvist}\ \emph {et~al.}(2009)\citenamefont
  {Broqvist}, \citenamefont {Alkauskas},\ and\ \citenamefont
  {Pasquarello}}]{Broqvist09}%
  \BibitemOpen
  \bibfield  {author} {\bibinfo {author} {\bibfnamefont {P.}~\bibnamefont
  {Broqvist}}, \bibinfo {author} {\bibfnamefont {A.}~\bibnamefont {Alkauskas}},
  \ and\ \bibinfo {author} {\bibfnamefont {A.}~\bibnamefont {Pasquarello}},\
  }\href {\doibase 10.1103/PhysRevB.80.085114} {\bibfield  {journal} {\bibinfo
  {journal} {Phys. Rev. B}\ }\textbf {\bibinfo {volume} {80}},\ \bibinfo
  {pages} {085114} (\bibinfo {year} {2009})}\BibitemShut {NoStop}%
\bibitem [{\citenamefont {Bang}\ \emph {et~al.}(2013)\citenamefont {Bang},
  \citenamefont {Sun}, \citenamefont {Abtew}, \citenamefont {Samanta},
  \citenamefont {Zhang},\ and\ \citenamefont {Zhang}}]{Bang13}%
  \BibitemOpen
  \bibfield  {author} {\bibinfo {author} {\bibfnamefont {J.}~\bibnamefont
  {Bang}}, \bibinfo {author} {\bibfnamefont {Y.~Y.}\ \bibnamefont {Sun}},
  \bibinfo {author} {\bibfnamefont {T.~A.}\ \bibnamefont {Abtew}}, \bibinfo
  {author} {\bibfnamefont {A.}~\bibnamefont {Samanta}}, \bibinfo {author}
  {\bibfnamefont {P.}~\bibnamefont {Zhang}}, \ and\ \bibinfo {author}
  {\bibfnamefont {S.~B.}\ \bibnamefont {Zhang}},\ }\href {\doibase
  10.1103/PhysRevB.88.035134} {\bibfield  {journal} {\bibinfo  {journal} {Phys.
  Rev. B}\ }\textbf {\bibinfo {volume} {88}},\ \bibinfo {pages} {035134}
  (\bibinfo {year} {2013})}\BibitemShut {NoStop}%
\bibitem [{\citenamefont {Kresse}\ and\ \citenamefont
  {Furthm\"uller}(1996)}]{Kresse96}%
  \BibitemOpen
  \bibfield  {author} {\bibinfo {author} {\bibfnamefont {G.}~\bibnamefont
  {Kresse}}\ and\ \bibinfo {author} {\bibfnamefont {J.}~\bibnamefont
  {Furthm\"uller}},\ }\href {\doibase 10.1103/PhysRevB.54.11169} {\bibfield
  {journal} {\bibinfo  {journal} {Phys. Rev. B}\ }\textbf {\bibinfo {volume}
  {54}},\ \bibinfo {pages} {11169} (\bibinfo {year} {1996})}\BibitemShut
  {NoStop}%
\bibitem [{\citenamefont {Freysoldt}\ \emph {et~al.}(2016)\citenamefont
  {Freysoldt}, \citenamefont {Lange}, \citenamefont {Neugebauer}, \citenamefont
  {Yan}, \citenamefont {Lyons}, \citenamefont {Janotti},\ and\ \citenamefont
  {Van~de Walle}}]{Freysoldt16}%
  \BibitemOpen
  \bibfield  {author} {\bibinfo {author} {\bibfnamefont {C.}~\bibnamefont
  {Freysoldt}}, \bibinfo {author} {\bibfnamefont {B.}~\bibnamefont {Lange}},
  \bibinfo {author} {\bibfnamefont {J.}~\bibnamefont {Neugebauer}}, \bibinfo
  {author} {\bibfnamefont {Q.}~\bibnamefont {Yan}}, \bibinfo {author}
  {\bibfnamefont {J.~L.}\ \bibnamefont {Lyons}}, \bibinfo {author}
  {\bibfnamefont {A.}~\bibnamefont {Janotti}}, \ and\ \bibinfo {author}
  {\bibfnamefont {C.~G.}\ \bibnamefont {Van~de Walle}},\ }\href {\doibase
  10.1103/PhysRevB.93.165206} {\bibfield  {journal} {\bibinfo  {journal} {Phys.
  Rev. B}\ }\textbf {\bibinfo {volume} {93}},\ \bibinfo {pages} {165206}
  (\bibinfo {year} {2016})}\BibitemShut {NoStop}%
\bibitem [{\citenamefont {Gr{\"u}neis}\ \emph {et~al.}(2014)\citenamefont
  {Gr{\"u}neis}, \citenamefont {Kresse}, \citenamefont {Hinuma},\ and\
  \citenamefont {Oba}}]{Gruneis14}%
  \BibitemOpen
  \bibfield  {author} {\bibinfo {author} {\bibfnamefont {A.}~\bibnamefont
  {Gr{\"u}neis}}, \bibinfo {author} {\bibfnamefont {G.}~\bibnamefont {Kresse}},
  \bibinfo {author} {\bibfnamefont {Y.}~\bibnamefont {Hinuma}}, \ and\ \bibinfo
  {author} {\bibfnamefont {F.}~\bibnamefont {Oba}},\ }\href {\doibase
  10.1103/PhysRevLett.112.096401} {\bibfield  {journal} {\bibinfo  {journal}
  {Phys. Rev. Lett.}\ }\textbf {\bibinfo {volume} {112}},\ \bibinfo {pages}
  {096401} (\bibinfo {year} {2014})}\BibitemShut {NoStop}%
\bibitem [{\citenamefont {Mills}\ and\ \citenamefont
  {J\'onsson}(1994)}]{Mills94}%
  \BibitemOpen
  \bibfield  {author} {\bibinfo {author} {\bibfnamefont {G.}~\bibnamefont
  {Mills}}\ and\ \bibinfo {author} {\bibfnamefont {H.}~\bibnamefont
  {J\'onsson}},\ }\href {\doibase 10.1103/PhysRevLett.72.1124} {\bibfield
  {journal} {\bibinfo  {journal} {Phys. Rev. Lett.}\ }\textbf {\bibinfo
  {volume} {72}},\ \bibinfo {pages} {1124} (\bibinfo {year}
  {1994})}\BibitemShut {NoStop}%
\bibitem [{\citenamefont {Selim}\ and\ \citenamefont
  {Kr{\"o}ger}(1977)}]{Selim77}%
  \BibitemOpen
  \bibfield  {author} {\bibinfo {author} {\bibfnamefont {F.}~\bibnamefont
  {Selim}}\ and\ \bibinfo {author} {\bibfnamefont {F.}~\bibnamefont
  {Kr{\"o}ger}},\ }\href {\doibase 10.1149/1.2133312} {\bibfield  {journal}
  {\bibinfo  {journal} {J. Electrochem. Soc.}\ }\textbf {\bibinfo {volume}
  {124}},\ \bibinfo {pages} {401} (\bibinfo {year} {1977})}\BibitemShut
  {NoStop}%
\bibitem [{Sup()}]{Supp_material}%
  \BibitemOpen
  \href@noop {} {}\bibinfo {note} {See Supplementary Material at [URL will be
  inserted by publisher] for the structural data used in this
  study.}\BibitemShut {Stop}%
\end{thebibliography}%

\end{document}